\documentclass[preprint, review, numbers, 1p, 12pt]{elsarticle}

 \usepackage[fleqn]{amsmath}
 \usepackage{amsthm, amssymb} 
\usepackage{arydshln} 
\usepackage{rotating}
\usepackage{setspace} 
\usepackage{lineno}

\setcitestyle{square}

\journal{Nuclear Instruments and Methods in Physics Research Section B: Beam Interactions with Materials and Atoms}

\newcommand\T{\rule{0pt}{2.6ex}}

\newcommand\TT{\rule{0pt}{3.5ex}}
\newcommand\BB{\rule[-2.2ex]{0pt}{0pt}}

\begin{document}

\begin{frontmatter}

\title{Ratio Estimation in SIMS Analysis}

\author[label1]{R. C. Ogliore\corref{cor1}}
\cortext[cor1]{Corresponding author}
\ead{ogliore@higp.hawaii.edu}
\author[label1]{G. R. Huss}
\author[label1]{K. Nagashima}
\address[label1]{Hawaii Institute of Geophysics and Planetology, University of Hawaii at Manoa, 1680 East-West Road, Honolulu, HI 96822, USA}

\begin{abstract}
The determination of an isotope ratio by secondary ion mass spectrometry (SIMS) traditionally involves averaging a number of ratios collected over the course of a measurement. We show that this method leads to an additive positive bias in the expectation value of the estimated ratio that is approximately equal to the true ratio divided by the counts of the denominator isotope of an individual ratio. This bias does not decrease as the number of ratios used in the average increases. By summing all counts in the numerator isotope, then dividing by the sum of counts in the denominator isotope, the estimated ratio is less biased: the bias is approximately equal to the ratio divided by the summed counts of the denominator isotope over the entire measurement. We propose a third ratio estimator (Beale's estimator) that can be used when the bias from the summed counts is unacceptably large for the hypothesis being tested. We derive expressions for the variance of these ratio estimators as well as the conditions under which they are normally distributed. Finally, we investigate a SIMS dataset showing the effects of ratio bias, and discuss proper ratio estimation for SIMS analysis.
\end{abstract}


\end{frontmatter}

\section{Introduction}
\label{intro}
Ratio estimation is broadly used in all scientific disciplines. Unfortunately, the mathematical issues associated with calculating meaningful ratios from experimental data are frequently ignored or misunderstood. In experiments dealing with blood sera in 1909, Greenwood \& White \cite{Greenwood:1909p4064} observed that a distribution of ratios (calculated as a mean of a number of individual ratios), randomly generated, was positively skewed. Karl Pearson explained the effect mathematically \cite{Pearson:1910p4930}. The statistical properties of ratio estimation were well-explored in the 1950s and 1960s (e.g. \cite{Hartley:1954p4025}, \cite{Tin:1965p4023}, \cite{Rao:1969p4022}), and these methods have been applied sparingly to various disciplines (e.g. \cite{Flueck:1976p3877}, \cite{VanKempen:2000p3771}). 

The crux of the problem of ratio estimation can be understood by considering a statistical variate $z$ defined in the range $0<z\leq\infty$ that has a probability p$(z_i)$ of taking on the value $z_i$. Assuming the expectation values of $z$ and $1/z$, $E\{z\}$ and $E\{1/z\}$, exist, we can calculate:
\begin{equation}
\label{equation:thecrux}
E\{z\}E\{1/z\}=\displaystyle\sum_{\textnormal{all }i} z_i\,\textnormal{p}(z_i)\displaystyle\sum_{\textnormal{all }k} \frac{1}{z_k}\,\textnormal{p}(z_k)
\end{equation}
If $z$ takes on just one value $z_j$ (such that the probability p$(z_j)=1$), this reduces to:
\begin{equation}
E\{z\}E\{1/z\}= \left(z_j\,\textnormal{p}(z_j)\right)\left( \frac{1}{z_j}\,\textnormal{p}(z_j)\right)=\textnormal{p}(z_j)^2=1
\end{equation}
If $z$ takes on more than one value ($z_1$, $z_2$,...), we expand the terms in Equation \ref{equation:thecrux}:  
\begin{equation}
E\{z\}E\{1/z\}= \left(z_1\,\textnormal{p}(z_1) + z_2\,\textnormal{p}(z_2) + ...\right)\left( \frac{1}{z_1}\,\textnormal{p}(z_1) + \frac{1}{z_2}\,\textnormal{p}(z_2) + ...\right)
\end{equation}
Collecting the squared terms and the cross terms for all possible values of $z$:
\begin{equation}
\label{equation:zoverz1}
E\{z\}E\{1/z\}= \left(\displaystyle\sum_{\textnormal{all } k} \textnormal{p}(z_k)^2\right)+\left( \displaystyle\sum_{m>n} \left(\frac{z_m}{z_n}+\frac{z_n}{z_m}\right)\textnormal{p}(z_m)\textnormal{p}(z_n) \right)
\end{equation}
We wish to compare this with $1$, which we can write as the square of the sum of all possible probabilities p($z_i$):
\begin{equation}
\label{equation:zoverz2}
1=\left(\displaystyle\sum_{\textnormal{all } i} \textnormal{p}(z_i)\right)\left(\displaystyle\sum_{\textnormal{all } i} \textnormal{p}(z_i)\right)= \left(\displaystyle\sum_{\textnormal{all } k} \textnormal{p}(z_k)^2\right)+\left( \displaystyle\sum_{m>n} 2\,\textnormal{p}(z_m)\textnormal{p}(z_n) \right)
\end{equation}
Using the fact:
\begin{equation}
\left(\frac{z_m}{z_n}+\frac{z_n}{z_m}\right) >2 \textnormal{\qquad for } z_m, z_n >0 \textnormal{ and } z_m \neq z_n
\end{equation}
we see that if $z$ that takes on more than one value, Equation \ref{equation:zoverz1} has a larger value than Equation \ref{equation:zoverz2} and:
\begin{equation}
E\{z\}E\{1/z\}>1
\end{equation}
Therefore we have shown that if $z$ is not single-valued:
\begin{equation}
E\{1/z\}>1/{E\{z\}}
\end{equation}
as also given in \cite{Kendall:1977p4158}. It follows that if $z$ is an unbiased estimator of some quantity $\theta$, $1/z$ cannot be an unbiased estimator of $1/\theta$ \cite{Kendall:1977p4158}. For two positive, statistically independent random variables $X$ and $Y$:
\begin{equation}
\label{equation:ratiobias}
E\{Y/X\}=E\{Y\}E\{1/X\}>E\{Y\}/E\{X\}
\end{equation}
showing that this ratio estimator is biased: an experiment attempting to measure $E\{Y\}/E\{X\}$ using $E\{Y/X\}$ will have an expectation value larger than the true value.

The current methods for determining isotope ratios in natural samples by secondary-ion mass spectrometry (SIMS) have their roots in other types of mass spectrometry.  In mass spectrometry, it is typically harder to determine the absolute abundance of an element than it is to determine the relative abundances of the isotopes of that element.  This naturally leads to reporting isotope data as ratios, eliminating the need to know the absolute abundance of any species.  A critical analytical problem is variation in the strength of the ion beam in the mass spectrometer.  If the mass spectrometer has only one collector, it is important to account for changes in ion-beam strength in order to get accurate isotope ratios.  Time interpolation is widely used to account for slow drifts in ion-beam strength.  Isotopes are measured in sequence (e.g., $^{24}$Mg, $^{25}$Mg, $^{26}$Mg, $^{24}$Mg, \ldots) such that each isotope samples the time variation in the ion-beam strength.  To correct for time drift, the signal for one isotope (e.g., $^{24}$Mg) is interpolated between two measurements to give the signal at the time when another isotope was measured.  The isotope ratio is then calculated for each cycle and the individual ratios are averaged to give the final result.  By collecting data as a series of ratios, one can also identify noise bursts or other problems that can affect the data, and cycles affected by the problem can be eliminated from the measurement.  If count rates are so low that an isotope gets zero counts in a cycle, calculating by-cycle ratios does not work (i.e., a ratio with a zero denominator is meaningless).  In this case, SIMS analysts typically add up all of the counts for each isotope from all of the cycles and calculate a single ratio for the measurement.  It is widely believed that the two methods, using the mean of the individual ratios to calculate the final result and using the ratio of the total counts for each isotope to calculate the final result, give the same answer.  But this is not the case.

In this paper, we attempt to understand how ratio-estimator bias affects SIMS analysis and provide a framework for calculating isotope ratios with less bias. We will look at three ratio estimators and investigate the first four statistical moments of each estimator: the mean, variance, skewness, and kurtosis. From these statistical moments we can determine the bias (from the first statistical moment, the expectation value), efficiency (from the second statistical moment, the variance), and approach to normality (from the third and fourth statistical moment, the skewness and kurtosis) of each ratio estimator. A specific example of how bias affects SIMS data will be discussed.

\section{First statistical moment of ratio estimators $r_1$ and $r_2$: expectation value}
\label{secexp}
We assume two populations $X$ and $Y$ from which $n$ subsamples $x$ and $y$ are randomly drawn. Our goal is to measure the ratio of the population means:
\begin{equation}
R=\frac{\overline{Y}}{\overline{X}}
\end{equation}
We can immediately think of two ways to estimate $R$ given $x$ and $y$. The ratio of the sample means:
\begin{equation}
r_1=\frac{\displaystyle \frac{1}{n}\displaystyle\sum_{i=1}^n y_i}{\displaystyle \frac{1}{n} \displaystyle\sum_{i=1}^n x_i}=\frac{\overline{y}}{\overline{x}}
\end{equation}
and the mean of the sample ratios:
\begin{equation}
r_2=\displaystyle \frac{1}{n}\displaystyle\sum_{i=1}^n \frac{y_i}{x_i} = \displaystyle \frac{1}{n}\displaystyle\sum_{i=1}^n \rho_i = \overline{\left(\frac{y}{x}\right)}
\end{equation}
Under most circumstances in mass spectrometry, $r_2$ is employed as the ratio estimator.

In this section we wish to calculate the first statistical moment (the expectation value) of each of these ratios ($E\{r_1\}$ and $E\{r_2\}$) to investigate the accuracy of the ratio estimators. 

Following \cite{Tin:1965p4023}, we can rewrite $r_1$ as:
\begin{equation}
r_1=R\left[1+\frac{\overline{y}-\overline{Y}}{\overline{Y}}\right]\left[1+\frac{\overline{x}-\overline{X}}{\overline{X}}\right]^{-1}
\label{r1expr}
\end{equation}
where $\overline{x}$ and $\overline{y}$ are the sample means, and $\overline{X}$ and $\overline{Y}$ are the population means. The factor 
\begin{equation}
\left[1+\frac{\overline{x}-\overline{X}}{\overline{X}}\right]^{-1}
\end{equation}
will converge as a geometric series if
\begin{equation}
\label{equation:convergenceconstraint}
\left|\frac{\overline{x}-\overline{X}}{\overline{X}}\right| < 1
\end{equation}
which is true under the assumption that $x$'s are positive (in SIMS data, this is equivalent to nonzero counts), and $n$ (the number of measurement cycles) is sufficiently large so that $\overline{x}<2\overline{X}$ (which is the case in any SIMS dataset where a reasonable measurement is sought).

With these assumptions we can expand Equation \ref{r1expr} as a geometric series:
\begin{equation}
\label{equation:r1expand}
\begin{split}
r_1=& R\left[1+\frac{\overline{y}-\overline{Y}}{\overline{Y}}\right]\left[1-\frac{(\overline{x}-\overline{X})}{\overline{X}}+\frac{(\overline{x}-\overline{X})^2}{\overline{X}^2}-\frac{(\overline{x}-\overline{X})^3}{\overline{X}^3} ... \right] \\
 =& R\left[1+\frac{\overline{y}-\overline{Y}}{\overline{Y}}-\frac{(\overline{x}-\overline{X})}{\overline{X}}-\frac{(\overline{x}-\overline{X})(\overline{y}-\overline{Y})}{\overline{X}\overline{Y}}+  \frac{(\overline{x}-\overline{X})^2}{\overline{X}^2}+\right. \\
& \left. \frac{(\overline{x}-\overline{X})^2(\overline{y}-\overline{Y})}{\overline{X}^2\overline{Y}}-\frac{(\overline{x}-\overline{X})^3}{\overline{X}^3}-\frac{(\overline{x}-\overline{X})^3(\overline{y}-\overline{Y})}{\overline{X}^3\overline{Y}}+\frac{(\overline{x}-\overline{X})^4}{\overline{X}^4}+...\right]
\end{split}
\end{equation}
At this point we assume that our SIMS measurements, the sampled isotope counts $x$ and $y$, follow two Poisson distributions with expected values equal to the population means $\overline{X}$ and $\overline{Y}$. The probability that $k$ counts of $x$ occur in a single sampling is $\overline{X}^ke^{-\overline{X}}/k!$; the probability that $k$ counts of $y$ occur in a single sampling is $\overline{Y}^ke^{-\overline{Y}}/k!$. For now we will assume $X$ and $Y$ are not independent.

Before we calculate $E\{r_1\}$, we note the following equalities (some of which are from \cite{Kendall:1977p4158}):
\begin{equation}
\label{equation:expectvals}
\begin{split}
&E\{(\overline{x}-\overline{X})\}=E\{(\overline{y}-\overline{Y})\}=0\\
&E\{(\overline{x}-\overline{X})^2\}=S^2(\overline{x})=\overline{X}/n\\
&E\{(\overline{x}-\overline{X})(\overline{y}-\overline{Y})\}=S(\overline{x},\overline{y})=S(x,y)/n\\
&E\{(\overline{x}-\overline{X})^2(\overline{y}-\overline{Y})\}=(S(x^2,y)-2\overline{X}S(x,y))/n^2\\
&E\{(\overline{x}-\overline{X})^3(\overline{y}-\overline{Y})\}=3\overline{X}S(x,y)/n^2+O(n^{-3})\\
&E\{(\overline{x}-\overline{X})^3\}=\overline{X}/n^2\\
&E\{(\overline{x}-\overline{X})^4\}=3\overline{X}^2/n^2+O(n^{-3})
\end{split}
\end{equation}
The population variance of $x$ is denoted by $S^2(x)$, and the population covariance of $x$ and $y$ is denoted by $S(x,y)$.

 We can now calculate $E\{r_1\}$ by substituting these expressions into Equation \ref{equation:r1expand}. Keeping terms to order $1/n^2$ (i.e., $O(n^{-2})$):
\begin{equation}
\label{equation:r1expec}
E\{r_1\} \approx R\left[1+\frac{1}{n}\left(\frac{1}{\overline{X}}-\frac{S(x,y)}{\overline{X}\,\overline{Y}}\right)+\frac{1}{n^2}\left(\frac{2}{\overline{X}^2}-\frac{S(x,y)}{\overline{X}\,\overline{Y}}\left(2+\frac{3}{\overline{X}}\right)+\frac{S(x^2,y)}{\overline{X}^2\,\overline{Y}}\right)\right]
\end{equation}
For independent $X$ and $Y$ this reduces to:
\begin{equation}
\label{equation:r1expecind}
E\{r_1\} \approx R\left[1+\frac{1}{n\overline{X}}+\frac{2}{n^2\overline{X}^2}\right]
\end{equation}
This shows that $r_1$, the ratio of the means, is a biased estimator of the ratio, R, consistent with Equation \ref{equation:ratiobias}.

For $r_2$ we note:
\begin{equation}
\label{equation:r2i}
\begin{split}
E\{r_2\}&=E\left\{\frac{1}{n}\displaystyle\sum_{i=1}^n \rho_i\right\}=\frac{1}{n}\left(E\{\rho_1\}+E\{\rho_2\}+...+E\{\rho_n\} \right)\\
&=\frac{1}{n}(nE\{\rho_i\})=E\{\rho_i\}=E\left\{\frac{y_i}{x_i}\right\}
\end{split}
\end{equation}
That is, the expectation value of the mean of the ratios is equal to the expectation value of a single ratio.
We choose a ratio $\rho_i$ for which the inequality in Equation \ref{equation:convergenceconstraint} holds (in a reasonable SIMS sample set, there would be at least one ratio that satisfies this constraint), and then we can expand $r_2$ in a power series and obtain an equation like Equation \ref{equation:r1expand}, with individual $x_i$ and $y_i$ replacing the sample means $\overline{x}$ and $\overline{y}$. Then, to obtain analogous expressions for the expectation values given in Equation \ref{equation:expectvals}, we substitute $n=1$ and keep terms to $O(\overline{X}^{-2})$.  This yields the expectation value for $r_2$ analogous to Equation \ref{equation:r1expec}:
\begin{equation}
\label{equation:r2expec}
E\{r_2\} \approx R\left[1+\frac{1}{\overline{X}}+\frac{2}{\overline{X}^2}-\frac{3S(x,y)}{\overline{X}\,\overline{Y}}\right]
\end{equation}
 For independent $X$ and $Y$ this reduces to:
\begin{equation}
\label{equation:r2expecind}
E\{r_2\} \approx R\left[1+\frac{1}{\overline{X}}+\frac{2}{\overline{X}^2}\right]
\end{equation}
which shows that $r_2$, the mean of the ratios, is also a biased estimator of $R$, but with a bias that is independent of the number of cycles $n$. Because of this, the bias of $r_2$ is usually much larger than $r_1$. This arises from Equation \ref{equation:r2i}: the first-order bias term for $r_1$, which is proportional to the variance of the mean of the denominator counts for $n$ cycles, becomes proportional to the variance of an individual measurement $x_i$ for $r_2$. Since the variance of the mean of the counts over $n$ cycles is equal to the variance of an individual measurement divided by $n$, the first-order bias term of $r_1$ is a factor of $1/n$ smaller than the first-order bias term of $r_2$.

The important conclusion is that the expectation values of $r_1$ and $r_2$ are both strictly greater than $R$:
\begin{equation}
\begin{split}
E\{r_{1,2}\}>&\frac{E\{y\}}{E\{x\}}=\frac{\overline{Y}}{\overline{X}}=R
\end{split}
\end{equation}

The difference between the expectation value of the ratio estimators $r_1$ or $r_2$ and the actual ratio $R$ is the bias:
\begin{equation}
\label{equation:bs}
\begin{split}
B\{r_1\}=&E\{r_1\}-R \approx R\left[\frac{1}{n\overline{X}}+\frac{2}{n^2\overline{X}^2}\right]\\
B\{r_2\}=&E\{r_2\}-R \approx R\left[\frac{1}{\overline{X}}+\frac{2}{\overline{X}^2}\right]
\end{split}
\end{equation}
for independent $x$ and $y$.
The bias of $r_1$ decreases as the total number of denominator counts in the entire measurement $n\overline{X}$ increases, but the bias of $r_2$ only decreases as the expected number of counts \emph{per cycle} $\overline{X}$ increases, independent of the total number of cycles $n$.

Ratio estimators can be constructed with smaller bias than $r_1$. Tin \cite{Tin:1965p4023} looked at four ratio estimators and found that Beale's ratio estimator \cite{Beale:1962p4364} has the least bias for finite and infinite source populations with $x$ and $y$ following a bivariate normal distribution. Additionally,  Tin \cite{Tin:1965p4023} found that Beale's estimator performed well for small samples, e.g. $n=50$. Srivastva et al. \cite{Srivastva:1983} confirmed that Beale's estimator outperformed the standard estimator $r_1$ in a bivariate normal model. Monte Carlo simulations by Hutchison \cite{Hutchison:1971p4030} of various ratio estimators and Poisson-distributed random variables showed that Beale's estimator performed best for small $n$ and small $\overline{X}$. We choose Beale's estimator for these reasons and also because of its simple structure and accuracy compared to the other estimators on very small and/or unusual data sets \cite{Tin:1965p4023}.

Beale's ratio estimator for $R=\overline{Y}/\overline{X}$, assuming infinite source populations, from which $x$ and $y$ are drawn, is:
\begin{equation}
r_3=r_1\displaystyle \frac{\left(1+\displaystyle \frac{\textnormal{cov}(x,y)}{n \, \overline{x}\,\overline{y}}\right)}{\left(1+\displaystyle  \frac{\textnormal{var}(x)}{n \, \overline{x}^2}\right)}
\end{equation}
where $n$ is the number of sample pairs $x_i$ and $y_i$, and:
\begin{equation}
\textnormal{cov}(x,y)=\frac{1}{n-1}\displaystyle\sum_{i=1}^n \left(x_i-\overline{x}\right)\left( y_i-\overline{y} \right)
\end{equation}
\begin{equation}
\textnormal{var}(x)=\frac{1}{n-1}\displaystyle\sum_{i=1}^n \left(x_i-\overline{x}\right)^2
\end{equation}
are the sample covariance and sample variance, respectively.

The expectation value of $r_3$, analogous to Equations  \ref{equation:r1expec} and \ref{equation:r2expec}, is computed by Tin \cite{Tin:1965p4023}. For our Poisson-distributed $x$ and $y$ and infinite source population, we find that Beale's estimator is unbiased to order $O(n^{-2})$. Beale's estimator, therefore, appears to perform much better than the standard ratio estimators $r_1$ and $r_2$ for our circumstances. The order $O(n^{-3})$ terms of $r_3$ assuming Poisson distributions are very complicated, but we note that the Poisson distribution is well-approximated by the normal (Gaussian) distribution for large mean values. The theoretical expectation value of $r_3$ assuming $x$ and $y$ are normally distributed only differs by 0.01\% when $\overline{X}=10$ from the expectation value assuming Poisson-distributed  $x$ and $y$.  We report the expectation value of $r_3$ for $\overline{X}$ and $\overline{Y}$ larger than 10 using a normal approximation to order $O(n^{-3})$ as deduced by Tin \cite{Tin:1965p4023} and Srivastva et al. \cite{Srivastva:1983}:
\begin{equation}
\label{equation:r3expecindnormal}
E\{r_3\} \approx R\left[1-\frac{2}{n^2\overline{X}}\left( \frac{1}{\overline{X}}-\frac{S(x,y)}{\overline{X}\overline{Y}} \right)\left(1+\frac{13}{2n}+\frac{8}{n\overline{X}} \right) \right]
\end{equation}
For normally distributed and independent $X$ and $Y$, the bias of $r_3$ is:
\begin{equation}
\label{equation:r3bias}
B\{r_3\}=E\{r_3\}-R \approx -R\left[\frac{2}{n^2\overline{X}^2}\left(1+\frac{13}{2n}+\frac{8}{n\overline{X}} \right)  \right]
\end{equation}
Unlike $r_1$ and $r_2$, the bias of $r_3$ is negative. 

\subsection{Computer simulations of the expectation values of $r_1$, $r_2$, and $r_3$}
\label{section:computersimulationsE}
The bias in these ratio estimators can be shown to approximately agree with the above equations by simulating a large number of experiments computationally. We calculate $r_1$, $r_2$, and $r_3$ using Poisson-distributed counts randomly generated by \texttt{poissrnd} in \texttt{MATLAB} version 7.10.0. For simplicity, we assume $R=\overline{Y}/\overline{X}=1$ for all computer simulations (the results are the same for all values of $R$). For each $\overline{X}$ and number of cycles $n=10,100,1000$, we generate $10^9$, $10^8$, and $10^7$ sets of simulated counts of $x$ and $y$ drawn from their associated Poisson distributions. For each of these sets, we calculate the ratio estimator $r_1$, $r_2$, and $r_3$.  We use the maximum likelihood estimator \texttt{mle} in \texttt{MATLAB} to determine the mean of the $10^9$, $10^8$, and $10^7$ sets (for a given $\overline{X}$ and $n$) and 95\% confidence interval of the mean. We compare these simulated values to the derived equations for the biases $B\{r_{1,2,3}\}$.
 
\begin{figure}[h]
       \centering
        \includegraphics[width=\textwidth]{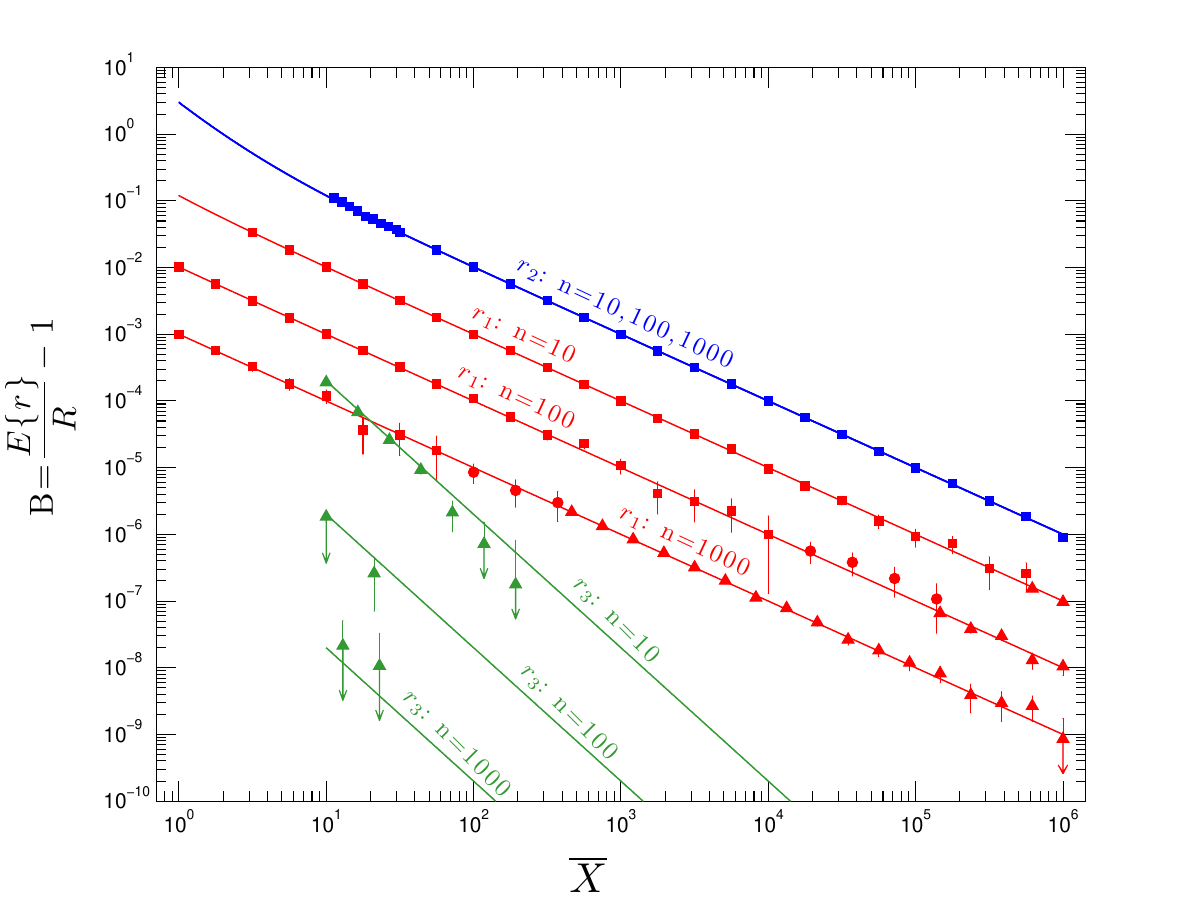}
        \caption{The calculated bias of $r_1$, $r_2$, and $r_3$ (Equations \ref{equation:bs} and \ref{equation:r3bias}) as a function of counts in the denominator for $n$=10, 100, and 1000 cycles are shown by curves. (The bias of estimator $r_2$ is independent of the number of samples $n$.) The computer simulations for Poisson-distributed $x$ and $y$ are given for each ratio estimator and $n$ by squares, the 95\% confidence interval of the bias are given by vertical lines when the uncertainty is larger than the size of the marker. Circles indicate computations where $x$ and $y$ are normally distributed, triangles indicate computations where $\overline{x}$, and $\overline{y}$, as well as sample variances and covariances, were drawn from a normal distribution.  Contributions from terms $O(\overline{X}^{-3})$ and smaller are evident for $B\{r_2\}$ near $\overline{X}=10$ where simulation points are slightly more biased than the line of calculated bias. For small $\overline{X}$, calculation of $r_2$ results in divide-by-zero, so those values are excluded. Since the bias of $r_3$ is negative, the opposite of $B\{r_3\}$ is plotted.}
        \label{figure:RatioBs}
 \end{figure}

For moderately large $n$ and $\overline{X}$, we approximate the Poisson distribution by a normal distribution, which greatly speeds up the random-number generation in \texttt{MATLAB}. To estimate the smallest biases, such as those for $r_3$, we make a further assumption that the means of the counts follow a normal distribution, as do the sample variances and covariances of these counts \cite{Tin:1965p4023}. The computer simulations closely follow the theoretical values, as shown in Figure \ref{figure:RatioBs}. 

\subsection{Time-varying count rate in SIMS data}
The count rate of individual isotopes in a SIMS measurement can vary by large amounts over the course of a measurement. One contributing factor to this phenomenon is variation in primary beam intensity. In addition, as more primary ions become implanted into the sample, the ionization efficiency increases, and the measured count rate in the electron multiplier or Faraday cup increases. If these types of experimental variations are large compared to the statistical variance of the data, the isotope counts show a strong sample covariance. However, the statistical behavior of the data actually reflects a time-varying population mean: $\overline{X}$ and $\overline{Y}$ are no longer constant, but are time-dependent, and each subsample $x_i$ and $y_i$ samples from $\overline{X}(t)$ and $\overline{Y}(t)$ over the $n$ cycles of the measurement.

We model this phenomenon with computer simulations by assuming the count rate changes (over a range from $\sim$10\% to several hundred percent) for the numerator and denominator over the course of the measurement following a sigmoidal function (continuously increasing) or various sinusoidal functions (increasing and decreasing). Our simulations show that the expectation value of $r_1$ and $r_3$ for these types of measurements should be calculated by substituting the measurement-time average of $\overline{X}(t)$ for $\overline{X}$ in Equation \ref{equation:r1expecind} and Equation \ref{equation:r3expecindnormal}. The analogous expectation value of $r_2$ is computed differently: the time-varying expectation value is calculated from Equation \ref{equation:r2expecind} by substituting $\overline{X}(t)$ for $\overline{X}$, and then the expectation value over the entire measurement is deduced by computing the average of this varying expectation value. The variances of $r_1$, $r_2$, and $r_3$ can be calculated in a similar way using the equations of Section \ref{varsec}. Using these methods, the statistical properties of a measurement with significant time-dependence can be understood.

\section{Second statistical moment of ratio estimators  $r_1$,  $r_2$, $r_3$: variance}
\label{varsec}
We wish to measure a ratio (biased, as it is) as precisely as possible given a fixed number of cycles and isotope counts per cycle. The most efficient ratio estimator will have the lowest statistical variance, given by the second statistical moment of the ratio estimator.

The variance of $r_1$, the ratio of the means, is:
\begin{equation}
V\{r_1\}= E\left\{\left(r_1-E\{r_1\}\right)^2\right\}=E\{r_1^2\}-E\{r_1\}^2
\end{equation}
Expanding Equations \ref{equation:r1expand} and \ref{equation:r1expec}, while substituting the expressions in Equation \ref{equation:expectvals}, and keeping terms of $O(n^{-2})$):
\begin{equation}
\begin{split}
\label{equation:vr1}
V\{r_1\} \approx &R^2\left[\frac{1}{n}\left(\frac{1}{\overline{X}}+\frac{1}{\overline{Y}}-\frac{2S(x,y)}{\overline{X}\,\overline{Y}}\right)+\right.\\
&\frac{1}{n^2}\left(\frac{6}{\overline{X}^2}+\frac{3}{\overline{X}\,\overline{Y}}+S(x,y)\left(\frac{4}{\overline{Y}^2}-\frac{8}{\overline{X}\,\overline{Y}}-\frac{16}{\overline{X}^2\overline{Y}}+\frac{5S(x,y)}{\overline{X}^2\overline{Y}^2}\right)+\right.\\
&\left.\left.\frac{4S(x^2,y)}{\overline{X}^2\overline{Y}}-\frac{2S(x,y^2)}{\overline{X}\overline{Y}^2}\right)\right]
\end{split}
\end{equation}
which in the case of independent $X$ and $Y$ reduces to:
\begin{equation}
\label{equation:r1var}
V\{r_1\} \approx R^2\left[\frac{1}{n}\left(\frac{1}{\overline{X}}+\frac{1}{\overline{Y}}\right)+\frac{1}{n^2}\left(\frac{6}{\overline{X}^2}+\frac{3}{\overline{X}\,\overline{Y}}\right)\right]
\end{equation}

Since $r_2$ is the mean of $n$ ratios, we can write its variance, assuming the individual ratios $y_i/x_i$ are uncorrelated, as:
\begin{equation}
V\{r_2\}=V\left\{\overline{\left(\frac{y}{x}\right)}\right\}=\frac{1}{n^2}\,V\left\{ \displaystyle\sum_{i=1}^n \frac{y_i}{x_i} \right\}=\frac{1}{n^2}\,\displaystyle\sum_{i=1}^n V\left\{\frac{y_i}{x_i} \right\}=\frac{1}{n}\,V\left\{\frac{y_j}{x_j} \right\}
\label{equation:vr2}
\end{equation}
If  the individual ratios $y_i/x_i$ are not independent:
\begin{equation}
V\{r_2\}=\left(\frac{1}{n}+\frac{\left(n-1\right)\rho}{n}\right)\,V\left\{\frac{y_i}{x_i} \right\}
\end{equation}
where $\rho$ is the average covariance between individual ratios. However, in the application of SIMS measurements, the individual ratios will in general be uncorrelated.
We substitute Equation \ref{equation:vr1} with $n=1$ into Equation \ref{equation:vr2} and keep terms to $O(\overline{X}^{-2})$ which yields an approximation for the variance of $r_2$:
\begin{equation}
\label{equation:vr22}
V\{r_2\} \approx \frac{R^2}{n}\left[\frac{1}{\overline{X}}+\frac{1}{\overline{Y}}-\frac{10S(x,y)}{\overline{X}\,\overline{Y}}+\frac{6}{\overline{X}^2}+\frac{3}{\overline{X}\,\overline{Y}}+\frac{4S(x,y)}{\overline{Y}^2}\right]
\end{equation}
In the case of independent $X$ and $Y$, this reduces to:
\begin{equation}
\label{equation:r2var}
V\{r_2\} \approx \frac{R^2}{n}\left[\frac{1}{\overline{X}}+\frac{1}{\overline{Y}}+\frac{6}{\overline{X}^2}+\frac{3}{\overline{X}\,\overline{Y}}\right]
\end{equation}

For Beale's estimator $r_3$, we assume the general case where $x$ and $y$ are Poisson-distributed. The variance of $r_3$ to order $O(n^{-2})$ as derived by Tin \cite{Tin:1965p4023} is:
\begin{equation}
\label{equation:vr3}
V\{r_3\} \approx R^2\left[\frac{1}{n}\left(\frac{1}{\overline{X}}+\frac{1}{\overline{Y}}-\frac{2S(x,y)}{\overline{X}\,\overline{Y}}\right)+\frac{1}{n^2}\left(\frac{2}{\overline{X}^2}-\frac{4S(x,y)}{\overline{X}^2\overline{Y}}+\frac{S(x,y)^2}{\overline{X}^2\,\overline{Y}^2}+\frac{1}{\overline{X}\,\overline{Y}}\right)\right]
\end{equation}
which in the case of independent $X$ and $Y$ reduces to:
\begin{equation}
\label{equation:r3var}
V\{r_3\} \approx R^2\left[\frac{1}{n}\left(\frac{1}{\overline{X}}+\frac{1}{\overline{Y}}\right)+\frac{1}{n^2}\left(\frac{2}{\overline{X}^2}+\frac{1}{\overline{X}\,\overline{Y}}\right)\right]
\end{equation}
For a large number of counts, the $O(n^{-1})$ approximation shows that all three ratio estimators $r_1$, $r_2$, $r_3$ have equal variance:
\begin{equation}
\label{equation:approxvar}
V\{r_{1,2,3}\} \approx R^2\left[\frac{1}{n}\left(\frac{1}{\overline{X}}+\frac{1}{\overline{Y}}-\frac{2S(x,y)}{\overline{X}\,\overline{Y}}\right)\right]
\end{equation}
which is the expected and familiar result \cite{Bevington:2003p4717}.

\subsection{Computer simulations of the variances of $r_1$, $r_2$, and $r_3$}
\label{section:computersimulationsV}

\begin{figure}
       \centering
        \includegraphics[width=\textwidth]{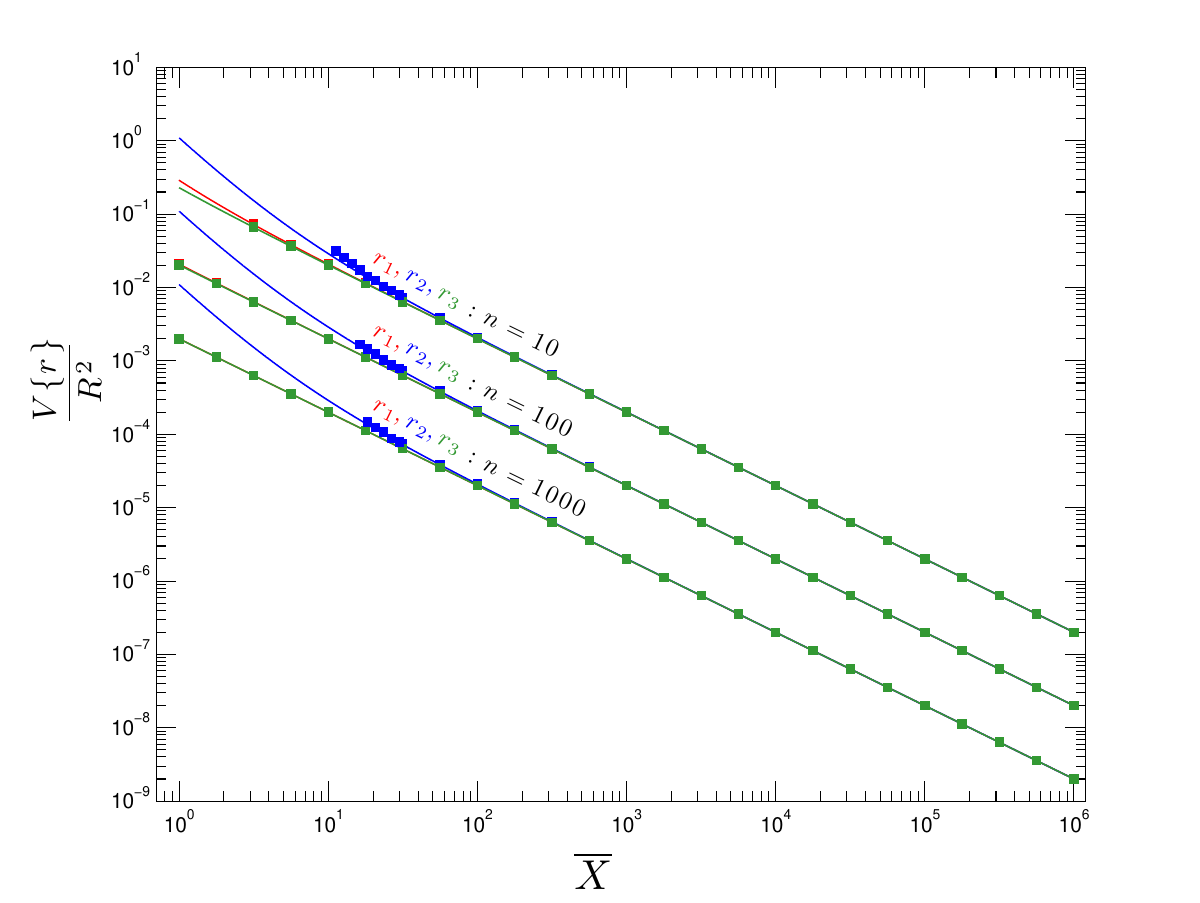}
        \caption{The calculated variance of $r_1$, $r_2$, and $r_3$ (Equations \ref{equation:r1var}, \ref{equation:r2var}, \ref{equation:r3var}) divided by $R$$^2$ as a function of counts in the denominator for 10, 100, and 1000 cycles are shown by curves. The computer simulations for Poisson-distributed $x$ and $y$ are given for each ratio estimator and $n$ by squares, the 95\% confidence interval of the variance is smaller than the size of the marker in all cases. More simulations were run for tens of counts in $\overline{X}$ to show the deviation of variance in $r_2$ from $r_1$ and $r_3$. Contributions from terms $O(\overline{X}^{-3})$ and smaller are evident for $V\{r_2\}$/$R$$^2$  near $\overline{X}=10$ and $n$ = 10 where the simulation points show slightly higher variance than the line of calculated variance. For $\overline{X}$ smaller than about 10, calculation of $r_2$ results in divide-by-zero, so those values are excluded from the plot. The calculated variance and computer simulations for $r_1$ are very close to $r_3$ and are thus obscured in this plot.}
        \label{figure:RatioVs}
 \end{figure}

Similarly, the variances $V\{r_{1,2,3}\}$ for the derived equations and computer simulations (calculated with $10^6$ sets) are shown in Figure \ref{figure:RatioVs}.
 For more than a few tens of counts per cycle, the variance follows the familiar expression given in Equation \ref{equation:approxvar}. The variance of the mean of the ratios, $r_2$, is substantially more than this for less than a few tens of counts per cycle. If a 1$\sigma$ uncertainty for $r_2$ is computed from the square root of the approximation given in Equation \ref{equation:approxvar}, this value will significantly underestimate the actual uncertainty of the $r_2$ ratio estimator. The variance of the ratio of means, $r_1$, is very slightly larger than Equation \ref{equation:approxvar} for very low counts.
 
 \section{Third and fourth statistical moment of ratio estimators of $r_1$,  $r_2$, $r_3$: skewness and kurtosis}
 The third and fourth statistical moments, the skewness ($\gamma_1$) and kurtosis ($\gamma_2$), determine if a random variate is approximately normally distributed. Assuming normally distributed $x$ and $y$, following Tin \cite{Tin:1965p4023}, to first order in $1/n$ the skewness of the ratio estimators $r_1$ and $r_3$ are:
 \begin{equation}
  \label{equation:r1skewness}
\gamma_1\{r_1\}= \frac{E\{\left(r_1-E\{r_1\}\right)^3\}}{\left(V\{r_1\}\right)^{3/2}}\approx
\end{equation}
\begin{equation*}
\left(\frac{\overline{Y}\Omega}{\left(n\overline{X}\,\overline{Y}^2\Omega^2+\overline{X}^2\,\overline{Y} \right)^{1/2}}\right)\left( 6+\left( 44+\frac{1}{\overline{Y}\Omega^2/\overline{X}+1} \right)\frac{1}{n\overline{X}} \right)
 \end{equation*}

\begin{equation}
 \label{equation:r3skewness}
\gamma_1\{r_3\}= \frac{E\{\left(r_3-E\{r_3\}\right)^3\}}{\left(V\{r_3\}\right)^{3/2}}\approx
\end{equation}
\begin{equation*}
\left(\frac{\overline{Y}\Omega}{\left(n\overline{X}\,\overline{Y}^2\Omega^2+\overline{X}^2\,\overline{Y} \right)^{1/2}}\right)\left( 6+\left( 26-\frac{1}{\overline{Y}\Omega^2/\overline{X}+1} \right)\frac{1}{n\overline{X}} \right)
 \end{equation*}

with:
\begin{equation*}
\Omega=\left(1-\overline{X}S(x,y)\right)
\end{equation*}

The skewness of $r_2$ can be found (analogous to our derivation of the variance of $r_2$) by setting $n=1$ in the expression for $\gamma_1\{r_1\}$ and then multiplying this expression by $1/\sqrt{n}$

 \begin{equation}
 \label{equation:r2skewness}
 \gamma_1\{r_2\} = \frac{E\{\left(r_2-E\{r_2\}\right)^3\}}{\left(V\{r_2\}\right)^{3/2}} \approx
\end{equation}
\begin{equation*}
\left(\frac{\overline{Y}\Omega}{\left(n\overline{X}\,\overline{Y}^2\Omega^2+n\overline{X}^2\,\overline{Y} \right)^{1/2}}\right)\left( 6+\left( 44+\frac{1}{\overline{Y}\Omega^2/\overline{X}+1} \right)\frac{1}{\overline{X}} \right)
 \end{equation*}

However, this expression underestimates the skewness of $r_2$ for $\overline{X}$ less than a few hundred counts. This is because there are more terms in the right-hand factor of Equation \ref{equation:r2skewness} (of order $1/\overline{X}^2$, $1/\overline{X}^3$,...) that increase the skewness of $r_2$ quickly for small $\overline{X}$. Computer simulations show that the $1/\overline{X}^2$ term is important for predicting the skewness of $r_2$ down to tens of counts in $\overline{X}$.

We calculate the kurtosis of the ratio estimators, again following Tin \cite{Tin:1965p4023}, and find that similar to the second and third statistical moments (variance and skewness), the kurtosis of $r_2$ is very close to the kurtosis of $r_1$ and $r_3$ down to a few tens of counts in $\overline{X}$. Below that, the kurtosis of $r_2$ increases extremely rapidly for $n=1000,100,10$ as shown by computer simulations. We do not report details of the kurtosis calculations or simulations here as they are similar to the results for skewness of the ratio estimators.

\subsection{Computer simulations of the skewness of $r_1$, $r_2$, and $r_3$}
\begin{figure}[h]
       \centering
        \includegraphics[width=\textwidth]{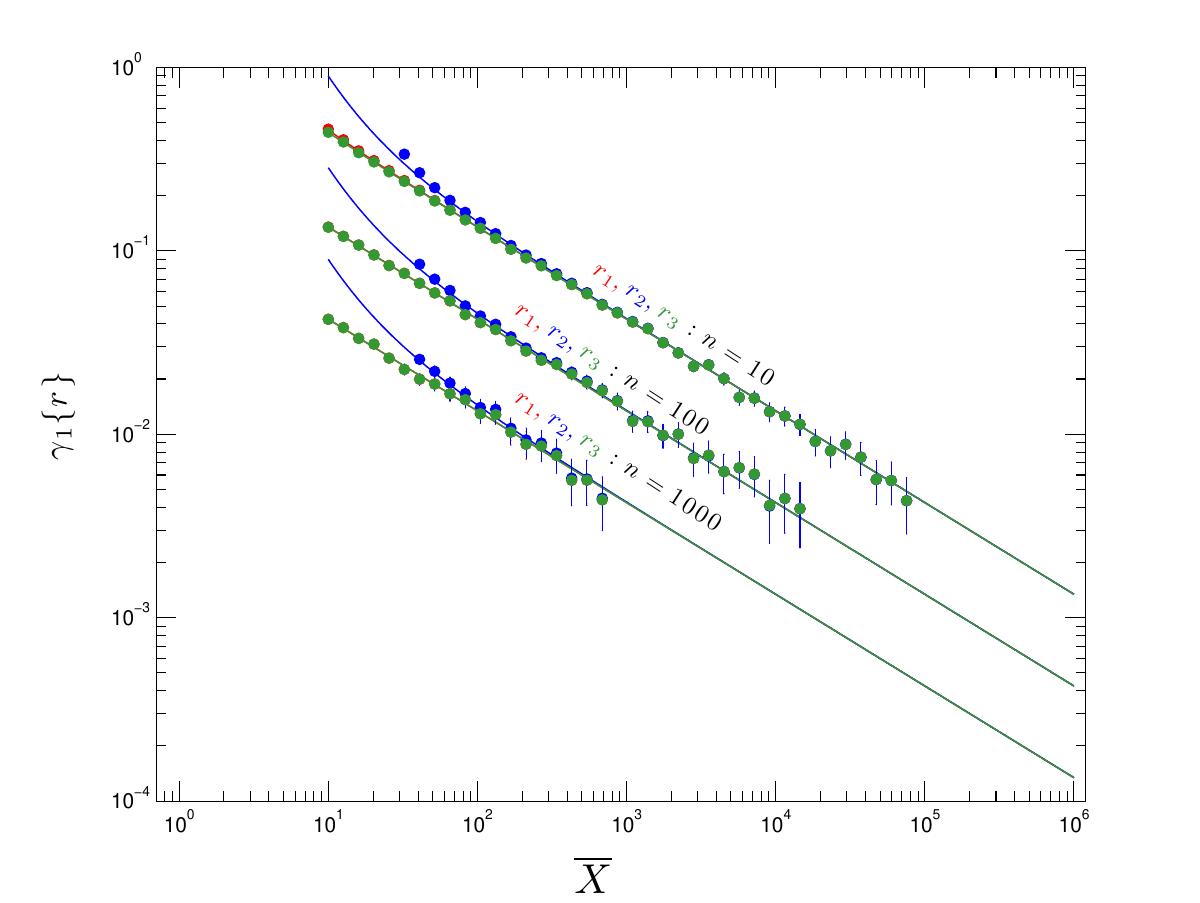}
        \caption{The calculated skewness of $r_1$, $r_2$, and $r_3$ (Equations \ref{equation:r1skewness},  \ref{equation:r2skewness}, and \ref{equation:r3skewness}) as a function of counts in the denominator $\overline{X}$ (equal to counts in the numerator $\overline{Y}$ for these calculations) for 10, 100, and 1000 samples are given by curves.  An approximation for the $1/\overline{X}^2$ term is included for the calculated skewness of $r_2$. The computer simulations for normally distributed $x$ and $y$ (so $\overline{X} \geq 10$) are given for each ratio estimator and $n$ by circles, the 95\% confidence interval of the skewnesses are given by error bars for some values of $r_2$, the uncertainties for $r_1$ and $r_3$ are always smaller than the size of the marker. For small $\overline{X}$, calculation of $r_2$ results in divide-by-zero, so those values are excluded.}
        \label{figure:RatioSs}
 \end{figure}

A normal distribution of $x$ and $y$ was assumed. Simulated random counts were created in \texttt{MATLAB} as for the simulations of expectation value and variance (Sections \ref{section:computersimulationsE} and \ref{section:computersimulationsV}), and the skewness and kurtosis of $10^4$ sets of ratios for each ratio estimator $r_1$, $r_2$, and $r_3$ were computed by \texttt{MATLAB}'s \texttt{skewness} and \texttt{kurtosis} functions. The mean and its 95\% confidence interval of 1000 kurtosis and skewness values were computed using \texttt{MATLAB}'s \texttt{mle} function. The results of the computer simulations for skewness of the three ratio estimators are shown in Figure \ref{figure:RatioSs}.

\subsection{Approach to normality of $r_1$, $r_2$, and $r_3$}
Ratio estimators in SIMS analysis are typically treated as variates following a normal distribution. Such a treatment greatly simplifies the task of calculating the statistical properties of a measurement, such as its statistical uncertainty. However, if the underlying distribution of the variate is not normal, it is not valid to treat it as such. Here we investigate under what conditions of the average number of counts per cycle ($\overline{X}$) and number of cycles ($n$) this assumption is valid for the three ratio estimators $r_1$, $r_2$, and $r_3$.

\begin{figure}[h]
       \centering
        \includegraphics[width=\textwidth]{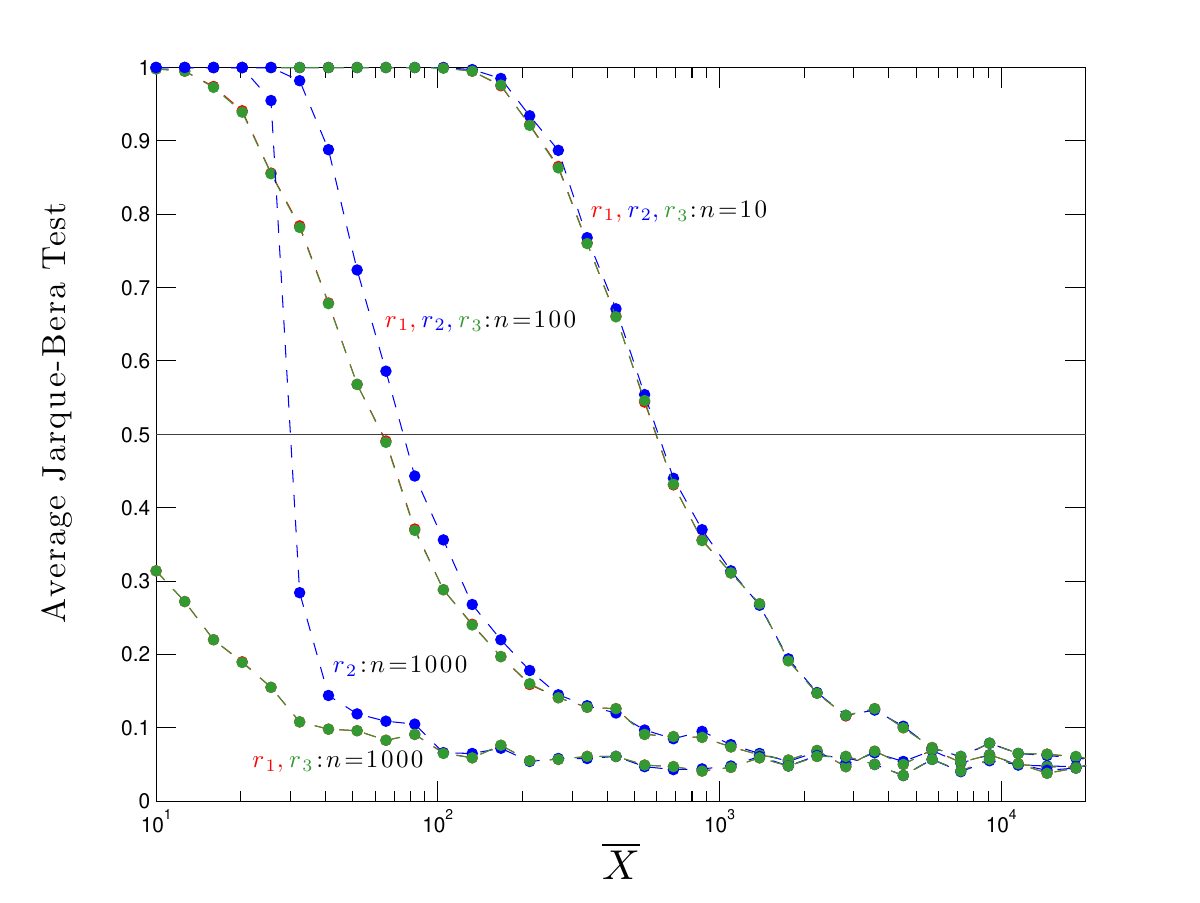}
        \caption{The average Jarque--Bera test for 1000 simulations of $r_1$, $r_2$, and $r_3$, with Poisson-distributed counts, as a function of average counts in the denominator $\overline{X}$. A value of zero means the null hypothesis ``the data is normally distributed" cannot be rejected at the 5\% significance level, a value of one means this hypothesis can be rejected at the 5\% significance level. The $\overline{X}$ value at which the curve becomes less than 0.5 is the number of counts we require for a given ratio estimator (with a given $n$) to be considered normally distributed. Values for $r_1$ are similar to values for $r_3$ so are mostly obscured.}.
        \label{figure:jbtest}
 \end{figure}

The Jarque--Bera hypothesis test uses the sample kurtosis and skewness to determine the departure from normality of a given distribution \cite{Jarque:1980p4541}. The test statistic is:
\begin{equation}
\textnormal{JB}=\frac{\eta}{6}\left(  \gamma_1^2 + \frac{\left( \gamma_2-3 \right)^2}{4}  \right)
\end{equation}
where $\eta$ is the sample size, $\gamma_1$ is the skewness and $\gamma_2$ is the kurtosis. We employ \texttt{MATLAB}'s \texttt{jbtest} function to determine when a ratio estimator ($r_1$, $r_2$, or $r_3$) with a given $\overline{X}$ and $n$ is normally distributed.  If there is less than a 5\% chance that the data is normally distributed, \texttt{jbtest} returns one, otherwise it returns zero. For $\eta>2000$ the statistic follows a $\chi^2$ distribution, for smaller $\eta$, \texttt{jbtest} uses results from Monte-Carlo simulations. We performed the Jarque--Bera test 1000 times on each set of $\eta=10^4$ ratios of $r_1$, $r_2$, and $r_3$ for a given $\overline{X}$ and $n$. From these 1000 tests, we calculated the mean of the Jarque--Bera test; results are shown in Figure \ref{figure:jbtest}.

Where the Jarque--Bera test returns zero, the distribution is normal. We assign $\overline{X}_{norm}$ to the number of counts per cycle where the average of the 1000 Jarque--Bera tests falls below 0.5 in our simulations. At this number of counts per cycle for a given number of cycles $n$, the ratio estimator can be considered normal, and standard normal statistics can be applied. The $\overline{X}_{norm}$ values for the three ratio estimators and $n=10,100,1000$ are given in Table \ref{xnormal}.

\begin{table}
\begin{tabular}{l c c}
\hline\hline  \TT \BB
$r$ & $n$ & $\overline{X}_{norm}$  \\  [0.5ex]
\hline
\T  $r_1$ &10 & 599 \\
$r_2$ &10 & 611 \\
$r_3$ &10 & 601 \\
\hdashline
$r_1$ &100 & 64 \\
$r_2$ &100 & 76 \\
$r_3$ &100 & 64 \\
\hdashline
$r_1$ &1000 & $<$10 \\
$r_2$ &1000 & 30 \\
$r_3$ &1000 & $<$10 \\
\hline
\end{tabular}
\caption{The minimum values required for normality of the three ratio estimators for number of cycles = 10, 100, 1000 as determined by computer simulations and the Jarque--Bera test.}
\label{xnormal}
\end{table}

\section{An example analysis of SIMS data using ratio estimators $r_1$, $r_2$, and $r_3$}
It is important to note that the biases derived in Section \ref{secexp} are for the expectation values of the ratio estimators: it is not true that for any given measurement $r_2 > r_1 > r_3$, only that the long-run average (if the measurement is performed an infinite number of times) of these ratio estimators will show this behavior. If the bias is small compared to the variance of a given measurement, the calculated ratio estimators have a reasonably high probability of not being ordered as $r_2>r_1>r_3$. Simply subtracting the bias from $r_2$ to reduce the bias in the calculated is ratio not a good strategy: the bias depends on the population mean $\overline{X}$ which is estimated in an unbiased way by $\overline{x}$. However $\overline{x}$ has a variance of $\overline{X}/n$ which means the estimation of the bias is uncertain, and the experimenter is better off using a ratio estimator with lower bias to begin with.

\begin{figure}[h]
       \centering
        \includegraphics[width=\textwidth]{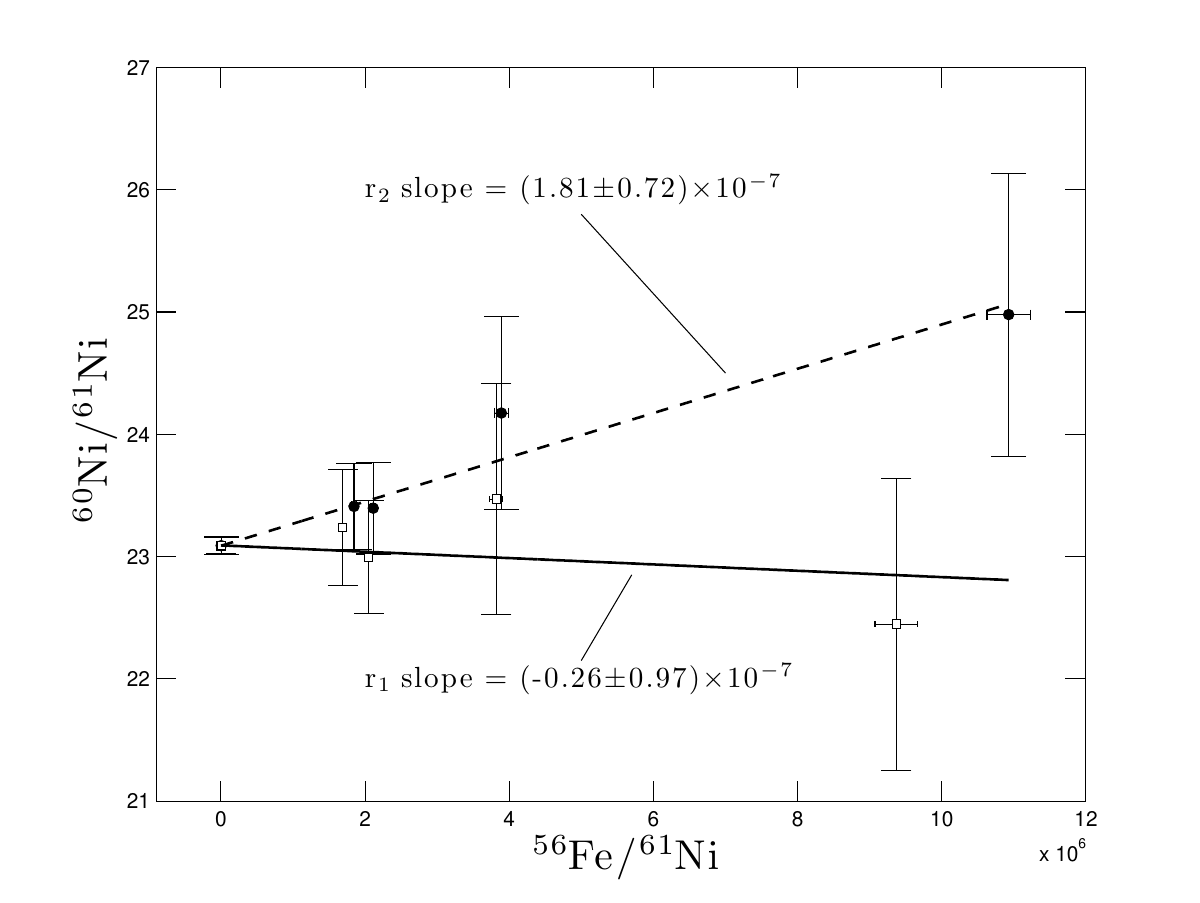}
        \caption{Fe--Ni isotopic measurements of troilite (FeS) in the Krymka LL3.1 ordinary chondrite made by a Cameca ims 6f secondary ion mass spectrometer (SIMS) at Arizona State University \cite{Tachibana:2003p4554}. Shown are two ratio estimators: $r_2$ (filled circles; data from \cite{Tachibana:2003p4554}) and $r_1$ (unfilled squares). Lines are fit to the data points using weighted total least-squares, the slope of which is equal to the inferred initial $^{60}$Fe/$^{56}$Fe in the troilite. A significant bias is associated with analyzing the data using the mean of ratios ($r_2$), resulting in a false detection of extinct $^{60}$Fe. This false detection disappears when a less-biased ratio estimator, $r_1$, is used on the same data. Isotopes were measured for 100 or 200 cycles ($n$) with 15--100 counts per cycle of $^{61}$Ni ($\overline{X}$).}
        \label{figure:fe60}
 \end{figure}

As an example of the dangers of using the ratio estimator $r_2$, we look at secondary ion mass spectrometry (SIMS) isotope measurements of troilite (FeS) in the LL3.1 ordinary chondrite Krymka. The presence of extinct short-lived radionuclides in meteoritic material can be explained by production due to energetic particle spallation or contribution from a nearby stellar source, such as a supernova. However, $^{60}$Fe is not efficiently produced by spallation, so the presence of its daughter $^{60}$Ni would be evidence of a stellar source. The Krymka meteorite has experienced little metamorphism to disturb the Fe--Ni isotopic system. When $^{60}$Ni/$^{61}$Ni is plotted against $^{56}$Fe/$^{61}$Ni, the SIMS measurements will fall on a line (an isochron), with slope equal to the initial abundance of $^{60}$Fe/$^{56}$Fe. Measurements of the Fe--Ni system in Krymka troilite were acquired on a Cameca ims 6f at Arizona State University. The ratios $^{60}$Ni/$^{61}$Ni and $^{56}$Fe/$^{61}$Ni were initially calculated using $r_2$, the mean of ratios, for 100 or 200 cycles with 15--100 counts per cycle of $^{61}$Ni ($\overline{X}$). Using $r_2$, the inferred initial abundance of $^{60}$Fe/$^{56}$Fe in troilite was (1.0--1.8)$\times$10$^{-7}$ \cite{Tachibana:2003p4554}. When these ratios are calculated using $r_1$, the ratio of means, this detection goes away, and the initial abundance of $^{60}$Fe/$^{56}$Fe in the Krymka troilite grains is no longer resolved from zero (see Figure \ref{figure:fe60}).

\section{Proper ratio calculations in mass spectrometry}
We have shown that the ratio estimator $r_2$, the mean of individually measured ratios, is poorly behaved in its first four statistical moments, especially with low counts.  Most importantly, its expectation value is heavily biased and this bias is independent of the number of cycles over which the mean of the ratios is calculated. The bias of $r_2$ can be as much as several percent for tens of counts in the denominator (reference) isotope. The variance of $r_2$ can be significantly worse than other ratio estimators for tens of isotope counts, resulting in much larger uncertainties on the measured ratio. Additionally, $r_2$ cannot be treated as a normally distributed variate for $\sim$30 counts per cycle unless $\sim$1000 cycles are measured, a factor of $\sim$3 more cycles than are needed for the two other ratio estimators.

The bias of $r_2$ is strictly positive; it cannot be thought of as another source of random error. It simply results in a calculated ratio higher than the actual source ratio. In many experiments, calculating $r_2$ instead of $r_1$ or $r_3$ is experimentally easier because instrumental conditions change over the course of the measurement.  Time interpolation of the ratios is a standard method of dealing with this problem.  However, one can time-interpolate the ion signals without calculating individual ratios for each cycle.  For example, if one wishes to calculate $r_1$, the ratio of the mean count rates, one can interpolate the signal for the denominator such that each measurement is modeled to have occurred at the same time as the measurement of the numerator.  The mean count rates can then be determined for the measured signal in the numerator and the calculated signal in the denominator, and the final ratio $r_1$ can be calculated.

In experiments where the ratio estimator $r_2$, the mean of ratios, must be used instead of $r_1$, the ratio of means, the effect of ratio bias on the final result of the measurement can be understood in terms of the relative bias of the ratio divided by the relative standard deviation of the ratio. For $r_2$, the relative bias of the ratio is independent of the number of cycles $n$, but the relative statistical standard deviation decreases as $n$ increases. If the expected bias is much smaller than the statistical uncertainty of the measurement, the bias can be ignored. We define the relative bias of the ratio divided by the relative standard deviation (the square root of the variance) for $r_2$. Using Equations \ref{equation:bs} and \ref{equation:r2var}, and keeping only first order terms:

 \begin{equation}
 \label{equation:bias_sigma}
\frac{B\{r_2\}/R}{\sqrt{V\{r_2\}}/R}\approx\frac{\frac{1}{\overline{X}}}{\sqrt{\frac{1}{n}\left(\frac{1}{\overline{X}}+\frac{1}{\overline{Y}}\right)}}=\sqrt{\frac{nR}{\overline{X}+\overline{Y}}}
\end{equation}

For the ratio bias associated with using $r_2$ to be insignificant, we require this quantity to be small. For example, if we require the bias to be less than 10\% of the statistical standard deviation of the measurement, we can deduce an upper bound on how many cycles $n$ can be used to calculate $r_2$:
 \begin{equation}
 \label{equation:bias_sigma2}
n < \frac{\left(\overline{X}+\overline{Y}\right)}{100\,R}
\end{equation}
That is, for the ratio bias associated with using $r_2$ (the mean of ratios) to be insignificant relative to the statistical uncertainty of the measurement, the number of cycles ($n$) must be smaller than the sum of expected counts ($\overline{X}$+$\overline{Y}$) divided by 100 times the expected ratio ($R$).

In multi-collector inductively coupled plasma mass spectrometry (MC-ICP-MS), count rates are typically higher than SIMS and, correspondingly, the measured ratios are usually more precise. Isotopes of Pb are measured using MC-ICP-MS with a precision of $\sim$50 ppm or better (e.g. \cite{Baker:2004p4949}). The ratio estimator $r_2$ is typically used in ICP-MS. In the case of Pb, the counts of all the isotopes in the collectors are so high ($\overline{X},\overline{Y}\approx10^{9}$) that the ratio bias from calculating $r_2$ over $\sim$100 cycles is insignificant (Equations \ref{equation:bias_sigma} and \ref{equation:bias_sigma2}).


In studies of radioactive nuclides and their daughter isotopes using SIMS, the daughter isotope is often in the numerator, and a high calculated ratio is often a positive detection and of scientific interest. A significant positive bias from calculating the ratio as $r_2$ is therefore potentially misleading and should be avoided whenever possible. The final ratio should only be calculated as $r_2$ if the bias can be estimated to be sufficiently small compared to the statistical uncertainty of the measurement (Equations \ref{equation:bias_sigma} and \ref{equation:bias_sigma2}) and any additional systematic uncertainty resulting from the use of $r_1$ or $r_3$ is significant.

The ratio of means, $r_1$ is also biased but the bias tends to zero as the number of cycles $n$ goes to infinity. For certain high-accuracy measurements, the bias of $r_1$, the ratio of the means, may still be too large to test the hypothesis at hand with sufficiently high confidence. In this case, Beale's estimator $r_3$ should be used, which has $\sim$1\% or less of the bias of $r_1$, and similar variance and approach to normality.

\section{Acknowlegements}
The authors wish to thank S. J. Desch and  G. J. Wasserburg for helpful comments, as well as the reviewer for suggestions which improved this manuscript. This work was supported by NASA grant NNX07AM62G to GRH. This is HawaiÔi Institute of Geophysics and Planetology publication No. 1886 and School of Ocean and Earth Science and Technology publication No. 8140.

%


\doublespacing
\section{Bibliography}

\bibliographystyle{elsarticle-num}

\bibliography{papers_bibtex}

\begin{thebibliography}{10}
\expandafter\ifx\csname url\endcsname\relax
  \def\url#1{\texttt{#1}}\fi
\expandafter\ifx\csname urlprefix\endcsname\relax\def\urlprefix{URL }\fi
\expandafter\ifx\csname href\endcsname\relax
  \def\href#1#2{#2} \def\path#1{#1}\fi

\bibitem{Greenwood:1909p4064}
M.~Greenwood, J.~D.~C. White, A biometric study of phagocytosis with special
  reference to the ``opsonic index", Biometrika 6~(4) (1909) 376--401.

\bibitem{Pearson:1910p4930}
K.~Pearson, On the constants of index-distributions as deduced from the like
  constants for the components of the ratio, with special reference to the
  opsonic index, Biometrika 7~(4) (1910) 531--541.

\bibitem{Hartley:1954p4025}
H.~Hartley, A.~Ross, Unbiased ratio estimators, Nature 174 (1954) 270--271.

\bibitem{Tin:1965p4023}
M.~Tin, Comparison of some ratio estimators, Journal of the American
  Statistical Association 60~(309) (1965) 294--307.

\bibitem{Rao:1969p4022}
P.~Rao, Comparison of four ratio-type estimates under a model, Journal of the
  American Statistical Association 64~(326) (1969) 574--580.

\bibitem{Flueck:1976p3877}
J.~Flueck, B.~Holland, Ratio estimators and some inherent problems in their
  utilization., Journal of Applied Meteorology 15~(6) (1976) 535--543.

\bibitem{VanKempen:2000p3771}
G.~van Kempen, L.~van Vliet, Mean and variance of ratio estimators used in
  fluorescence ratio imaging, Cytometry Part A 39~(4) (2000) 300--305.

\bibitem{Kendall:1977p4158}
M.~{Kendall}, A.~{Stuart}, {The advanced theory of statistics. Vol.1:
  Distribution theory}, {Griffin}, 1977.

\bibitem{Beale:1962p4364}
E.~Beale, Some uses of computers in operational research, Industrielle
  Organisation 31 (1962) 51--52.

\bibitem{Srivastva:1983}
V.~Srivastva, T.~Dwivedi, Y.~Chaubey, S.~Bhatnagar, Finite sample properties of
  beale's ratio estimator, Communications in Statistics - Theory and Methods
  12~(15) (1983) 1795--1805.

\bibitem{Hutchison:1971p4030}
M.~Hutchison, A {M}onte {C}arlo comparison of some ratio estimators, Biometrika
  58~(2) (1971) 313.

\bibitem{Bevington:2003p4717}
P.~R. {Bevington}, D.~K. {Robinson}, {Data reduction and error analysis for the
  physical sciences}, {McGraw-Hill}, 2003.

\bibitem{Jarque:1980p4541}
C.~Jarque, A.~Bera, Efficient tests for normality, homoscedasticity, and serial
  independence of regression residuals, Economics Letters 6~(3) (1980)
  255--259.

\bibitem{Tachibana:2003p4554}
S.~Tachibana, G.~Huss, The initial abundance of $^{60}${F}e in the solar
  system, The Astrophysical Journal Letters 588 (2003) L41--L44.

\bibitem{Baker:2004p4949}
J.~Baker, D.~Peate, T.~Waight{\ldots}, Pb isotopic analysis of standards and
  samples using a $^{207}${P}b-$^{204}${P}b double spike and thallium to
  correct for mass bias with a double-focusing {M}{C}-{I}{C}{P}-{M}{S},
  Chemical Geology 211 (2004) 275--303.

\end{thebibliography}

\end{document}